\documentclass[]{ceurart}

\usepackage{multicol}

\usepackage{ifthen}
\newboolean{showcomments}
\setboolean{showcomments}{true}          
\ifthenelse{\boolean{showcomments}}
  {\newcommand{\nb}[2]{
    \fcolorbox{gray}{yellow}{\bfseries\scriptsize#1}
    {\scriptsize$\blacktriangleright${#2}$\blacktriangleleft$}
   } 
  }
  {\newcommand{\nb}[2]{}}

\usepackage{listings}
\usepackage{float}
\floatstyle{plain}
\newfloat{model}{hbt}{lop}
\floatname{model}{Listing}

\begin{document}

\copyrightyear{2025}
\copyrightclause{Copyright for this paper by its authors.
  Use permitted under Creative Commons License Attribution 4.0
  International (CC BY 4.0).}

\conference{}

\title{Towards Sustainability Model Cards}

\author[1]{Gwendal Jouneaux}[%
orcid=0000-0003-1158-9335,
email=gwendal.jouneaux@list.lu,
url=https://www.gwendal-jouneaux.fr/,
]

\author[1,2]{Jordi Cabot}[%
orcid=0000-0003-2418-2489,
email=jordi.cabot@list.lu,
]
\address[1]{Luxembourg Institute of Science and Technology, Esch-sur-Alzette, Luxembourg}
\address[2]{University of Luxembourg, Esch-sur-Alzette, Luxembourg}

\begin{abstract}
    The growth of machine learning (ML) models and associated datasets triggers a consequent dramatic increase in energy costs for the use and training of these models.
    In the current context of environmental awareness and global sustainability concerns involving ICT, Green AI is becoming an important research topic. Initiatives like the AI Energy Score Ratings are a good example. Nevertheless, these benchmarking attempts are still to be integrated with existing work on Quality Models and Service-Level Agreements common in other, more mature, ICT subfields. This limits the (automatic) analysis of this model energy descriptions and their use in (semi)automatic model comparison, selection, and certification processes. 
    
    We aim to leverage the concept of quality models and merge it with existing ML model reporting initiatives and Green/Frugal AI proposals to formalize a Sustainable Quality Model for AI/ML models. As a first step, we propose a new Domain-Specific Language to precisely define the sustainability aspects of an ML model (including the energy costs for its different tasks). This information can then be exported as an extended version of the well-known Model Cards initiative while, at the same time, being formal enough to be input of any other model description automatic process.
\end{abstract}


\maketitle

\section{Introduction}
\label{sec:intro}
The large adoption of AI technologies, increase in model complexity, and dataset size has led to a dramatic rise in the computational power required to run and train the models and their overall energy cost and sustainability impact.
In the last few years, the question of the carbon footprint of AI models became a priority inside the research community.
The seminal paper from Strubell \textit{et al.} ~\cite{Strubell2019Energy} analyzed the carbon impact for training four state-of-the-art NLP models.
The resulting conclusion is that the carbon footprint for training and using AI models should be reduced.

While efforts to benchmark, monitor and fine-tune AI models in the context of Green AI has been made~\cite{verdecchia2023systematic}, this information is usually not readily available for model users.
Approaches such as Model Cards~\cite{mitchell2019model} provide detailed information to model users, such as performance metrics or ethical concerns.
Yet, few approaches bridge the gap and present sustainability information.
The most notable is the recent AI Energy Score~\cite{EnergyScore} proposed by HuggingFace.
However, this energy score only provides information on inference energy consumption, disregarding carbon emission, water usage or training impact.

To address the aforementioned issue, this paper proposes the definition of Sustainability Model Cards.
Sustainability Model Cards complement the already existing Model Cards~\cite{mitchell2019model} with the additional concern of sustainability of AI models.
We propose a domain-specific language (DSL) to precisely define these cards and pave the way for future automated use of this information.

The rest of the paper is organized as follows. Section~\ref{sec:related} reviews the state of the art on the modeling of quality models for AI, formalism to describe AI models and related artifacts, and existing approaches to make model users aware of sustainability aspects of the models.
Sections~\ref{sec:SMC} and Section~\ref{sec:DSL} detail the Sustainability Model Cards and associated DSL, respectively.
Finally, Section~\ref{sec:discussion} discusses future work regarding Sustainability Model Cards in the form of a research roadmap, while Section~\ref{sec:conclusion} concludes this paper.

\section{State of the art}
\label{sec:related}
Quality Assurances (QA) and related quality models have been identified as a challenge for current AI software research~\cite{felderer2021quality}.
In the past years, the research community proposed multiple quality models~\cite{gezici2022systematic, ali2022systematic}.
However, most of those models do not present a full picture of AI software quality.
Among the twenty-nine papers studied by Gezici \textit{et al.}~\cite{gezici2022systematic}, only three discuss sustainability as a relevant quality aspect~\cite{pons2019priority, siebert2020towards, horkoff2019non}.

In addition to quality models, other formalisms to describe AI models or related artifacts have been developed.
For datasets, Dataset Cards~\cite{DatasetCards} allows defining standard information such as provenance, authorship, license, and tasks for which the dataset is suitable.
DescribeML~\cite{giner2023domain} additionally describes social concerns potentially leading to bias and provides a dedicated language and tool support for its specification.
Finally, Croissant~\cite{akhtar2024croissant} uses a JSON notation that is both human-readable and compatible with existing tools and frameworks, providing additional interoperability, portability, and discoverability to datasets.
For AI models, Mitchell \textit{et al.} proposed Model Cards~\cite{mitchell2019model} used for model reporting.
Model Cards describe the AI model in terms of intended use, data, performances and ethical considerations, among other things.
However, none of those approaches allows the description of the sustainability concern.

On the other hand, approaches such as the ones from Hugging Face try to specifically assess and describe sustainability aspects of machine learning models.
They extended the Model Cards approach with sustainability data~\cite{ModelCardsCO2} such as cloud provider location, training time, hardware, and estimated carbon emissions.
They also created the AI Energy Score Ratings~\cite{EnergyScore} to evaluate the energy cost when using a model.
This method compute the average inference cost (in Wh) over one thousand requests.

However, these approaches neither provide a formal description that could be used to check the syntactic and semantic correctness of the card information nor an easy way to automatically process such information as part of a MLOps pipeline that should take into account energy concerns~\cite{cruz2025greeningaienabledsystemssoftware}. 
Furthermore, while carbon emissions of the training phase and energy consumption of the inference task are important metrics for model selection, our DSL covers a larger variety of energy-related information, such as the water consumption or counter measures taken by the platforms (and platform providers) to counter such energy impact. Finally, the use of DSL technology facilitates reusing many other existing tools in the DSL realm to simplify the creation of new functionality (e.g. verbalization of the card information) around this new Sustainability Model Card ecosystem.

\section{Sustainability Model Cards}
\label{sec:SMC}
To address the aforementioned problem, we propose Sustainability Model Cards, inspired by the well-known  Model Cards~\cite{mitchell2019model} concept. 
In what follows, we describe the dimensions that are part of our Sustainability definition based on an analysis of the relevant literature in this field.
More specifically, we group Sustainability information of a ML model into four main sections: Metadata, Training, Inference, and Platform. 

\begin{table}[!h]
\centering
\resizebox{\textwidth}{!}{%
\begin{tabular}{|llll|}
\hline
\multicolumn{4}{|c|}{\textbf{Sustainability Model Card}\rule[-10pt]{0cm}{25pt}}                                                                                                                                                          \\
\multicolumn{1}{|c}{\textbf{Metadata}} & \multicolumn{1}{c}{\textbf{Training}} & \multicolumn{1}{c}{\begin{tabular}[t]{@{}c@{}}\textbf{Inference}\\\textit{(for each task)}\rule[-10pt]{0cm}{10pt}\end{tabular}} & \multicolumn{1}{c|}{\textbf{Platform}} \\
-- Name                                & -- Training Duration                  & -- Inference Type                                                                                              & -- Hardware Details                \\
-- Version                             & -- Energy Consumption                 & -- Energy Consumption                                                                                          & -- Platform Provider                   \\
-- Type                                & -- Carbon Emissions                   & -- Carbon Emissions                                                                                            & -- Platform Region                     \\
-- Provider                            & -- Water Consumption                  & -- Water Consumption                                                                                           & -- Carbon Offset Credit                \\
-- License                             & -- \textit{Platform}                  & -- \textit{Platform}                                                                                           & -- Energy Sources\rule[-10pt]{0cm}{10pt}                     \\ \hline
\end{tabular}%
}
\end{table}

The \textbf{Metadata} section contains information about the model itself.
This includes the identification of the model through a name and version, the type of the model (\textit{e.g.,} decision tree, CNN, regression), identification of the provider and license of the model.
This information allows establishing the link between the Sustainability Model Cards and already existing Model Cards for the same model to later do some combined analysis. 
The model type is particularly important, as studies such as Yu \textit{et al.} shows potential correlation between model type and inference energy consumption~\cite{yu2022energy}.

The \textbf{Training} section regroup information concerning the environmental impact of the training phase of the model. More important aspects are the energy consumption, carbon emission and water consumption resulting from this training phase.
While carbon emissions and energy consumption are already studied in the context of Green AI~\cite{Strubell2019Energy}, water consumption is often overlooked, yet it is still regarded as an important sustainability aspect for datacenters~\cite{ristic2015water}.
To complement this information, this section also includes a reference (represented italicized) to the definition of the platform used, and the time spent to train the model (Which could be used to infer some energy consumption a posteriori with a certain degree of approximation).

The \textbf{Inference} section allows defining the impact of the different inference tasks a model can be used for.
For each task supported by the model (\textit{e.g.,} text generation, text summarization), this section reports the inference task type, the average energy consumption, the average carbon emission, the average water consumption, and a reference (represented italicized) to the platform used to compute these estimations.

Finally, the \textbf{Platform} section describes the platforms used to train or execute the model. 
This description includes details about the hardware used, the region (\textit{e.g.,} Azure and EU-west), the carbon offset credit bought to compensate the carbon emissions (in percentage or \(kgCO_2eq\)) and the energy mix used for the training (the ratio of renewable and fossil energy used), as these aspects are useful when choosing a deployed model infrastructure. 

\section{Description of a DSL for ML Sustainability reporting}
\label{sec:DSL}
This section describes the proposed domain-specific language (DSL) to support the definition of Sustainability Model Cards, including all dimensions mentioned in the previous section. A DSL is a language specially designed to model apps for a certain domain (health, finance, or, such as in this case, AI).

Once described with our DSL, the Sustainability information can be automatically processed as part of a ML workflow or MLOps approach (\textit{e.g.,} to search for sustainable alternatives, compare the sustainability of equivalent models, analysis of a set of models with a energy perspective,..).

Each DSL has two main components: 
\begin{itemize}
    \item Abstract syntax: Describing the structure of the language and the way different language primitives can be combined, independently of any particular representation.
    \item 	Concrete syntax. Describes one or more specific notations for the language, covering the encoding and/or the visual/textual appearance of the elements in the abstract syntax
\end{itemize}

As we will see, our DSL uses a YAML based concrete syntax to facilitate the adoption of the language and its integration with the existing Model Cards of the Hugging Face repository.

\subsection{Abstract Syntax of the Language}

\begin{figure}[hbt]
    \centering
    \includegraphics[width=\textwidth]{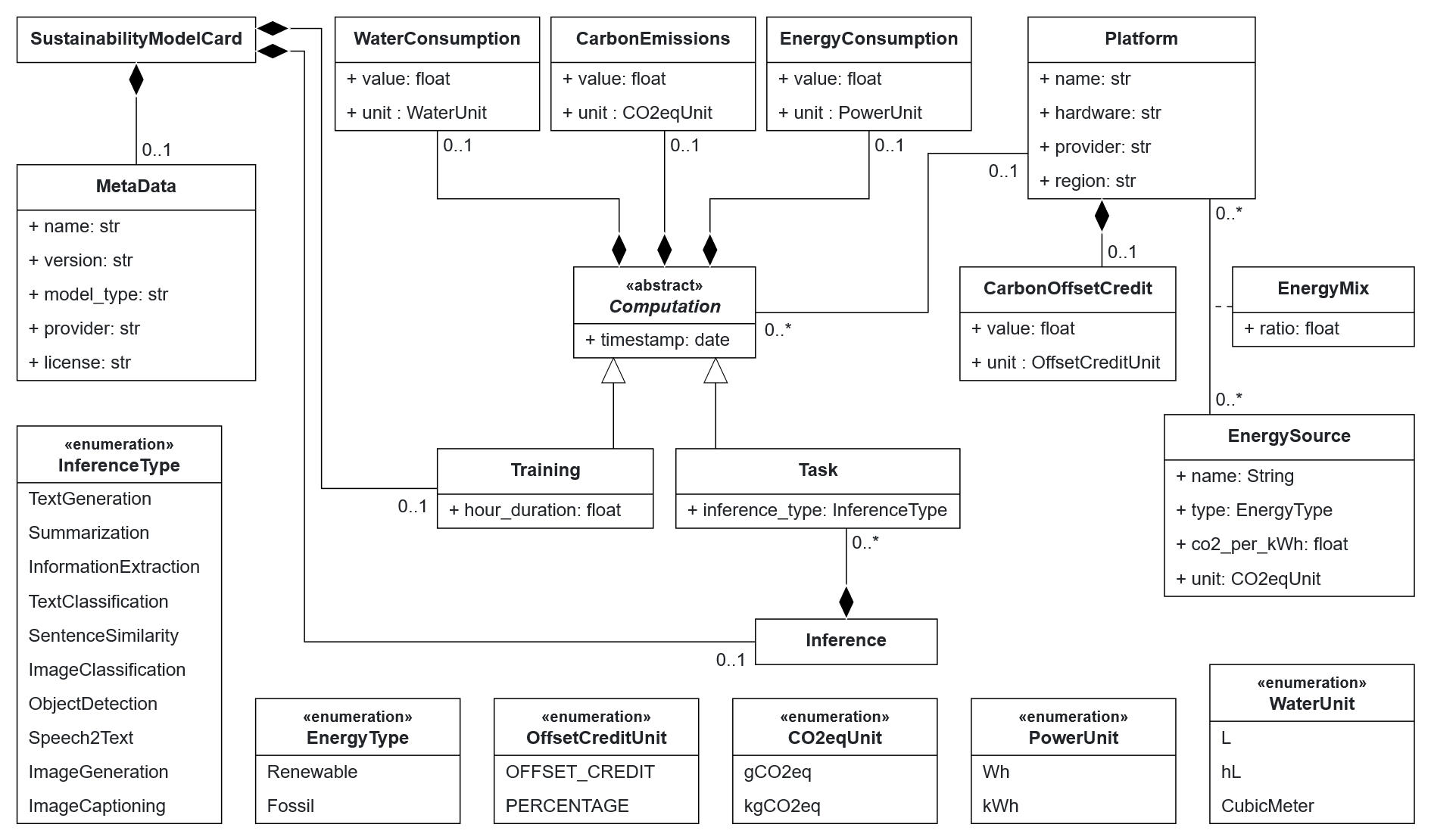}
    \caption{Metamodel of the Sustainability Model Cards DSL}
    \label{fig:Metamodel}
\end{figure}

Figure~\ref{fig:Metamodel} presents the abstract syntax of the proposed DSL in the form of a metamodel (i.e. the schema or grammar of the language expressed using an object-oriented perspective) structuring the language elements, the properties of every element and the possible relationships among them. 

Let's describe in more details the different elements of the language.
The \texttt{SustainabilityModelCard} class is the root concept representing the whole Sustainability Model Card.
This card is composed of three subcomponents: \texttt{MetaData}, \texttt{Training} and \texttt{Inference}.

\pagebreak
The \texttt{MetaData} class represents the metadata section of the card defining the name, version, model type, provider, and license as strings.

The \texttt{Training} class models the training section of the card and defines the duration of the training.
The \texttt{Inference} class encompasses all the inference tasks (represented by the \texttt{Task} class) addressed by the model.
Each \texttt{Task} defines the inference type, based on the list used in the AI Energy Score~\cite{EnergyScore}.
Both the \texttt{Training} and \texttt{Task} classes represent computations that have an environmental impact.
This is materialized through their inheritance from the \texttt{Computation} abstract class.
This class models the impact of the computation when executed on a given platform.
The impact is materialized through associations to water consumption, carbon emissions and energy consumption metrics and a timestamp denoting the moment of the measurement.
These three metrics are reified in their own class and are represented through a value and its associated unit.

In addition, the \texttt{Computation} class is associated to a \texttt{Platform} representing the infrastructure used to train the model or to compute the metrics, depending on the subclass instantiated.
The \texttt{Platform} class has a name, used to link it to its associated \texttt{Training} and/or \texttt{Task} in the concrete syntax, and details on the hardware, provider and compute region.
For a more fine-grained representation of the platform carbon impact, the \texttt{Platform} is associated to a \texttt{CarbonOffsetCredit} and a set of \texttt{EnergySource} --- containing the type of energy and carbon efficiency, representing its energy mix.

While \texttt{Training} and \texttt{Task} classes are mostly identical, we explicitly make a distinction based on their semantic difference and intent to have a more fine-grained description of the training phase, as presented in Section~\ref{sec:discussion}.

\subsection{Concrete Syntax of the Language}
The concrete syntax used to specify the Sustainability Model Card is based on YAML.
A YAML structure is represented as a set of key-value pairs, where the value can be of three types: Scalar, Sequence, and Mapping.
Scalar values are represented as a series of zero or more characters, Sequence values are represented as a list of values, and Mapping values are represented as a set of key-value pairs.

\pagebreak
To encode Sustainability Model Cards in this format, we defined a set of rules.
\begin{enumerate}
  \item Class instances are represented using the class name in snake case as key and a mapping as value
  \item Attributes are part of instances mapping and represented using the attribute name in snake case as key and a scalar as value
  \item Compositions are defined by having the composed class instance nested in the containing class mapping
  \item Multiplicity higher than one is managed using YAML sequence
  \item Simple associations are defined using the associated class name in snake case as key and the object name attribute as scalar value
  
\end{enumerate}

In addition, there are three special cases to these rules: \texttt{Platform} and \texttt{EnergySource}, \texttt{Inference}, and \texttt{EnergyMix}.
Even if platforms and energy sources are not direct components of the Sustainability Model Card, they are defined as a list contained in the card (See Listing \textit{platforms} and \textit{energy\_sources} lists).
As \texttt{Inference} only contains the list of tasks, the intermediate "\textit{task}" attribute containing the list is bypassed, making the YAML \textit{inference} section a sequence.
Finally, the \texttt{EnergyMix} association class is represented as a sequence of \texttt{EnergyMix} class instances containing the ratio as attribute and the \texttt{EnergySource} class instance.

\lstdefinestyle{mystyle}{
    keywordstyle=\color{magenta},
    numberstyle=\tiny\color{gray},
    basicstyle=\ttfamily\small,                   
    captionpos=b,                    
    keepspaces=true,                 
    numbers=left,                    
    numbersep=5pt
}
\begin{model}
\begin{center}
  \begin{minipage}[t]{0.49\textwidth}
\lstset{style=mystyle}
\begin{lstlisting}
sustainability_model_card:
  meta_data:
    name: Model Name
    version: 1.0.0
    model_type: LLM
    provider: Provider Name
    license: CC0
  platforms:
    - platform:
        name: Infrastructure
        hardware: GTX 1080 Ti
        provider: Microsoft Azure
        region: West Europe
        carbon_offset_credit:
          value: 100.0
          unit: PERCENTAGE
        energy_mix:
          - energy_mix:
              ratio: 100.0
              energy_source: Azure EU-W
  energy_sources:
    - energy_source:
        name: Azure EU-W
        type: Fossil
        co2_per_kWh: 0.57
        unit: kgCO2eq
\end{lstlisting}
  \end{minipage}
  \begin{minipage}[t]{0.49\textwidth}
  \lstset{style=mystyle}
\begin{lstlisting}[firstnumber=27]
  training:
    hour_duration: 100.0
    platform: Infrastructure
    carbon_emissions:
      value: 14.25
      unit: kgCO2eq
    energy_consumption:
      value: 25.0
      unit: kWh
    water_consumption:
      value: 57.5
      unit: L
    timestamp: 2025-01-02T09:00:00
  inference:
    - task:
        inference_type: TextGeneration
        platform: Infrastructure
        carbon_emissions:
          value: 7.12
          unit: gCO2eq
        energy_consumption:
          value: 12.3
          unit: Wh
        water_consumption:
          value: 0.02
          unit: L
        timestamp: 2025-01-21T09:00:00
\end{lstlisting}
  \end{minipage}
\end{center}
\caption{Syntax example using all the metamodel concepts}
\label{dsl-example}
\end{model}

As example, Listing~\ref{dsl-example} define the Sustainability Model Card for an LLM under Creative Commons license and describe the \textit{Infrastructure} platform defined as using a GTX 1080 Ti GPU and being provided by Microsoft Azure on the West Europe region.
The model is reported to have been trained for one hundred hours on the \textit{Infrastructure} platform.
The card also provides estimation for the text generation inference task based on what has been observed on the same \textit{Infrastructure} platform.

\subsection{Tool support}
To support the formal definition of Sustainability Model Cards, we provide a Python implementation of our DSL.
This implementation is composed of a validating parser and a set of classes implementing the metamodel, and is available in open-source on GitHub\footnote{Implementation: \url{https://www.gwendal-jouneaux.fr/SustainabilityModelCards-Parser}}.

To implement the metamodel, we relied on the BESSER Low-Code platform~\cite{alfonso2024building} to create the DSL metamodel and generate a Python implementation of it.
The Python classes generated can then be used to instantiate any model conforming to the specified metamodel, allowing its manipulation or creation by other tools.
For instance, BESSER offers both a language to specify and generate implementation of neural networks~\cite{daoudi2025modelling}, and a deployment language~\cite{ul2024extending}.
In the future, BESSER could use this infrastructure to automatically benchmark specified neural networks and generate their Sustainability Model Card as part of the design and execution process of the network.

In addition, we have implemented a parser validating and transforming the YAML description to model instances.
First, we use an existing YAML parser implementation, transforming the textual description into a manipulable Python object. 
Then, we traverse the object structure to assess the conformance of the structure to the metamodel.
These validation checks ensure: (1) the presence of units when required, (2) the correspondence of these units to the ones defined in the metamodel, (3)  the correspondence of the inference and energy types to the ones defined in the metamodel, and (4) that the values representing percentage are bound to the [0,1] interval.
Finally, the validated structure is transformed to a corresponding model instance using the metamodel classes.

\section{Research Roadmap}
\label{sec:discussion}
Our DSL is a first step towards a more ambitious goal towards the adoption, formalization, analysis and improvement of AI sustainability. In this section, we discuss a research roadmap and possible next steps for the evolution of the Sustainability Model Cards initiative, including a number of application scenarios. We hope to extend and prioritize this list based on discussions with the community. 

\textbf{Extending the coverage and granularity of the Sustainability DSL}.
When creating the Sustainability Model Cards and associated DSL, we aimed to be as complete as possible by making the union of sustainability concepts mentioned in the surveyed papers~\cite{Strubell2019Energy,strubell2020energy,ModelCardsCO2,EnergyScore,guldner2021exploration, yu2022energy,anthony2020carbontracker,garcia2019estimation,ristic2015water}.
Yet, more details could be added such as a more granular description of the training phase diving in the pre-training and fine-tuning part of the training, the hyperparameters values, or the dataset used.
Furthermore, the Sustainable AI research field is highly active and propose new frameworks and new metrics that should be included in Sustainability Model Cards along the way.
As expressed in the recent article of Cruz \textit{et al.}~\cite{cruz2025greeningaienabledsystemssoftware}, there is a need for standardized metrics in the evaluation of AI sustainability, and our framework should evolve to match (and possibly, influence) these upcoming metrics.

\textbf{Graphical notation}.
When using a language, different users prefers different syntaxes, also depending on their technical profile. For instance, less technical users tend to prefer more graphical notations. 
For this reason, we plan to extend the language with other concrete syntaxes, such as a graphical notation, and even a conversational-based one, for input and a verbalization in Controlled English as output to help all types of users to write and read sustainability cards.

\textbf{Tighter integration with Model Cards}. 
Sustainability Model Cards focus only on the sustainability aspect of AI models.
One of the ways to have a complete view of the model data would be through the integration of this card with other existing cards.
A first step in this direction has already been made, as our DSL concrete syntax has been built on YAML to integrate seamlessly with the existing Hugging Face model cards.
The next step in this direction would be to propose a formal description language integrating both cards, allowing automatic processing using all the available information.
Another direction would be to extend the sustainability concern with ethical concern to allow more transparency in the ethical and environmental sustainability impact of AI models~\cite{luccioni2025bridginggapintegratingethics}.

\pagebreak
\textbf{Analyzing impact on model users}. 
An additional aspect of these cards is the social aspeect.
The final choice of using the model with the most accuracy, the least carbon emissions, or something in the middle belongs to the model user. 
Conducting a user study on how they decide on a model, especially among models that offer similar features and performance, would allow assessing the impact of providing more sustainability data in the choice of model users. 

\textbf{Application on different scenarios}.
The precise models of the Sustainability Model Cards resulting from the use of our DSL allow for the automatic processing of the information reported in the cards.
This can be useful for a range of different scenarios, including many MLOps ones.
For instance, a first scenario envisioned is to perform automatic model selection based on the model environmental impact.
Another scenario would be to optimize model deployment based on an impact analysis using information on locations' energy providers and carbon efficiency, energy consumption of inference, and hardware.
Finally, the provided information could be used and monitored at runtime to enforce sustainability aware Service Level Agreements (SLA) that could be established between the users and the model providers as typically done for other types of IT services following existing quality models. 

\section{Conclusion}
\label{sec:conclusion}
In this paper, we have presented the Sustainability Model Cards, a new DSL allowing the description of AI models sustainability aspects including energy consumption, carbon emissions and water consumption.
This DSL enables the formal definition of this information and facilitates the automatic processing of sustainability information as part of a MLOps pipeline while still exporting it as an extension of the Models Card formalism for better readability and integration with similar initiatives. 

As further work, we plan to work on the aspects discussed as potential future roadmap for this initiative.

\bibliography{SustainabilityModelCard}

\end{document}